\documentclass[10pt]{article}
\usepackage{epsfig}
\usepackage{amssymb}
\usepackage{amsmath}

\usepackage{txfonts}
\usepackage{graphicx}
\usepackage{color}
\usepackage{colortab}
\usepackage{pstricks}
\usepackage{enumerate}
\usepackage[normalem]{ulem}

\usepackage{natbib}
\bibpunct{(}{)}{;}{a}{}{,} 
\definecolor{seda}{gray}{.83}

\newgray{lg}{0.85} 
\newgray{mg}{0.7} 

\begin{document}
\begin{center}
{{\bf\Large Multi-resonance orbital model\\applied to~high-frequency quasi-periodic oscillations

\vspace{1ex}
observed in Sgr\,A$^*$}}\\
\vspace{3ex}
 {\large Andrea Kotrlov\'a, Zden\v{e}k Stuchl\'{\i}k and Gabriel T\"{o}r\"{o}k}\\
\vspace{3ex}
{Institute of Physics, Faculty of Philosophy and Science, Silesian University in Opava, Bezru\v{c}ovo n\'{a}m. 13,
CZ-74601 Opava, Czech Republic\\
{\small e-mail: andrea.kotrlova@fpf.slu.cz}}
\end{center}

\vspace{3ex}
\begin{center}
{ABSTRACT}
\end{center}
\noindent{\small The multi-resonance orbital model of high-frequency quasi-periodic oscillations (HF~QPOs) enables precise determination of the~black hole dimensionless spin $a$ if observed set of oscillations demonstrates three (or more) commensurable frequencies. The black hole spin $a$ is related to the~frequency ratio only, while its mass $M$ is related to the~frequency magnitude. The model is applied to the~triple frequency set of HF~QPOs observed in Sgr\,A$^*$ source with frequency ratio $3:2:1$. Acceptable versions of the~multi-resonance model are determined by the~restrictions on the~Sgr\,A$^*$ supermassive black hole mass. Among the~best candidates the~version of strong resonances related to the~black hole ``magic'' spin $a=0.983$ belongs. However, the~version demonstrating the~best agreement with the~mass restrictions predicts spin $a=0.980$.}


\section{Introduction}

In the~black hole systems observed in both Galactic and extragalactic sources, strong gravity effects play a~crucial role in three phenomena related to the~accretion disk that is the~emitting source: the~spectral continuum, spectral profiled lines, and oscillations of the~disk.

Quasi-periodic oscillations (QPOs) of X-ray brightness had been observed at {low}-(Hz) and {high}-(kHz) frequencies in many Galactic low mass X-ray binaries (LMXBs) containing neutron~stars \citep[see, e.g.,][]{Kli:2000:ARASTRA:,Kli:2006:CompStelX-Ray:,Bar-Oli-Mil:2005:MONNR:,Bel-Men-Hom:2005:ASTRA:,Bel-Men-Hom:2007:MONNR:BriNSQPOCor} or black holes \citep[see, e.g.,][]{McCli-Rem:2004:CompactX-Sources:,Rem:2005:ASTRN:,Rem-McCli:2006:ARASTRA:}. Some of the~HF~QPOs are in the~kHz range and often come in pairs of the~upper and lower frequencies ($\nu_{\mathrm{U}}$, $\nu_{\mathrm{L}}$) of twin peaks in the~Fourier power spectra. Since the~peaks of high frequencies are close to the~orbital frequency of the~marginally stable circular orbit, representing the~inner edge of Keplerian disks orbiting black holes (or neutron~stars), the~strong gravity effects must be relevant in explaining HF~QPOs \citep{Abr-etal:2004:RAGtime4and5:CrossRef}.

The~resonance orbital model of HF~QPOs in black hole systems \citep{Ter-Abr-Klu:2005:ASTRA:QPOresmodel} is now partially supported by observations, in particular, when frequency ratio~3\,:\,2 ($2\nu_{\mathrm{U}} = 3\nu_{\mathrm{L}}$) is seen in twin peak QPOs in the~LMXBs containing black holes (microquasars), namely GRO~1655$-$40, XTE~1550$-$564, GRS~1915$+$105 \citep{Ter-Abr-Klu:2005:ASTRA:QPOresmodel}. Nevertheless, there is a~clear problem with explanation of the~3\,:\,2 frequency ratios in all three microquasars using a~resonance model with unique variant of the~twin oscillating modes, especially when the~limit on the~black hole spin given by the~spectral continuum fitting \citep{Rem-McCli:2006:ARASTRA:,McCli-etal:2011:CLAQG:} is used \citep{Tor-Kot-Sra-Stu:2011:,Ali-etal:2013:CQG,Ste-Gyu-Yaz:2012:arXiv}.

However, in the~case of the~GRS~1915$+$105 source the~frequency set of HF~QPOs is more complex -- in fact, at least five HF~QPOs were observed there \citep{Rem-McCli:2006:ARASTRA:}. Therefore, in this case we have to consider more complex models of HF~QPOs. We are able to explain the~complete observed frequency set in the~framework of the~extended resonant orbital model \citep{Stu-Sla-Tor:2007:ASTRA:,Stu-Sla-Ter:2006:ASTRA:Humpy} based on the~so-called Aschenbach effect \citep{Asch:2004:ASTRA:,Stu-etal:2004:PHYSR4:}. Another possibility is related to the~multi-resonance orbital model recently proposed in \citet{Stu-Kot-Tor:2013:ASTRA:MultiRez}.

The multi-resonance orbital model can be profoundly tested in one of the~most interesting and relevant cases related to the~observations in the~X-ray spectra of the~Galaxy centre supermassive black hole in the~source Sgr\,A$^*$ \citep{Abr-etal:2004:ASTRJ2L:,Asch:2004:ASTRA:,Ter:2005:ASTRA:}, where three frequency set with the~$3\,:\,2\,:1$ ratio has been observed. In the~multi-resonance orbital model of HF~QPOs, the~resonance conditions on the~frequency ratio of oscillations entering the~resonances imply specific radii of accretion disks where the~resonances can occur. For any specific version of the~orbital resonance model, the~resonant radii at thin, Keplerian accretion disks \citep[][]{Ter-Abr-Klu:2005:ASTRA:QPOresmodel} are given by the~parameters of the~central black hole, i.e., its mass $M$ and dimensionless spin $a$.\footnote{For toroidal disks \citep{Rez-Ahm-Mil:1991:MONNR:,Rez-etal:2003:MNRAS:,Bla-etal:2006:ASTRJ2:,Str-Sra:2009:CLAQG:EpiOscNonSleKerrBH,Mon-Zan:2012:MNRAS:}, slightly charged disks orbiting in combined strong gravitational and electromagnetic fields \citep{Bak-etal:2012:CQG:,Bak-etal:2010:CLAQG:MagIndNonGeo,Kov-Stu-Kar:2008:CLAQG:OffEqOrb}, or string loops oscillating in vicinity of the~black hole  horizon \citep{Stu-Kol:2012:PHYSR4:,Stu-Kol:2012:JCAP:}, additional parameters must enter the~resonance conditions.} The~frequencies of the~Keplerian disk oscillation modes can be expressed in terms of the~geodetical orbital and epicyclic frequencies of the~test particle motion in the~Kerr spacetime \citep{Ter-Stu:2005:ASTRA:}. When HF~QPOs are observed as a~three frequency set, the~multi-resonance orbital model can work in two ways. The first one, of strong resonant phenomena, assumes that all three oscillations in resonance occur at one fixed radius \citep{Stu-Kot-Tor:2008:ACTA:BHadmStrResPhen}. The second one, of duplex frequencies, assumes two radii where twin oscillations occur with one of the~frequencies being common \citep[see][for details]{Stu-Kot-Tor:2013:ASTRA:MultiRez}.

In the~framework of the~multi-resonance model triple sets of frequency ratios determine the~dimensionless black hole spin $a$ precisely, quite independently of the~black hole mass $M$, but not uniquely as several versions of the~model can enter the~play \citep{Stu-Kot-Tor:2013:ASTRA:MultiRez}. The mass parameter $M$ is then given by the~magnitude of the~observed frequencies. The precision of the~frequency measurement thus determines precision of the~mass parameter of the~black hole. We expect the~multi-resonance model will be applicable due to development of the~observational techniques \citep[e.g., the~planned LOFT observatory,][]{Fer-etal:2012:SPIE,Fer-etal:2012:ExA:}.

Here we apply the~multi-resonance orbital model to the~observations of three HF QPOs reported by Aschenbach \citep{Asch:2004:ASTRA:}, using the~precision of the~frequency measurements to give ranges of allowed values of the~Sgr\,A$^*$ central black hole mass for each of the~versions of the~multi-resonance model allowing for the~frequency ratio $3:2:1$. These mass ranges are compared with the~mass range given by the~motion of the~stars orbiting the~central black hole as presented in \citet{Gil-etal:2009:Mass-SgrA:} and acceptable versions of the~multi-resonance orbital model are thus found.

Of course, the~spin (and mass) estimates of the~Sgr\,A$^*$ black hole based on acceptable versions of the~multi-resonance model have to be compared with the~spin estimates based on the~accretion disk spectral continuum fitting \citep{McCli-etal:2011:CLAQG:,Don-Dav:2008:,Don-Gie-Kub:2007:,McCli-etal:2006:astro-ph/0606076:,Mid-etal:2006:MONNR:,Sha:2005::astro-ph/0508302} and profiled spectral lines \citep{Laor:1991:ASTRJ2:,Bao-Stu:1992:ASTRJ2:,Kar-Vok-Pol:1992:ASTRJ2L:,Fab-Min:2005:XSpectraKerr:Book,Zak:2003:,Cad-Cal-Fan:2003:MEMSA1:XrFeProf,Fan-etal:1997:,Cad-Cal:2005:MNRAS:,Zak-Rep:2006:,Sch-Stu:2009:GENRG2:ProEmRingBran,Mil-etal:2009:,Stu-Sch:2012:CQG:ObsPhenKerrSSp:} or by the~measurements of the~black hole (or superspinar) shadow \citep{Vir-Ell:2002:,Sch-Stu:2009:INTJMD:OpPheBraKerr,Stu-Sch:2010:CLAQG:AppKepDiOrKerrSSp,Eir:2012:arXiv}. In the~case of Sgr\,A$^*$ also the~orbital precession of some stars moving in close vicinity of the~central black hole could give interesting restriction on the~black hole spin \citep[]{Kra:2005:,Kra:2007:}.

\section{Orbital resonance model}

The~standard orbital resonance model \citep{Abr-Klu:2001:ASTRA:,Ter-Abr-Klu:2005:ASTRA:QPOresmodel,Ali-Gal:1981:GENRG2:} assumes non-linear resonance between oscillation modes of an~accretion disk orbiting a~central object, here considered to be a~rotating Kerr black hole.\footnote{We consider here only the~Kerr spacetime as the~standard description for rotating black holes, although alternatives have been discussed in the~same context \citep[see][]{Kot-Stu-Ter:2008:CLAQG:,Stu-Kot:2009:GENRG2:OrResDiBraKBH,Ali-Tal:2009:PHYSR4:RotBraBH,Rah-Abd-Ahm:2011:,Abd-Ahm-Hak:2011:PRD,Abd-Ahm:2010:PRD,Ali-etal:2013:CQG,Hor-Ger:2012:,Ste-Gyu-Yaz:2012:arXiv}.}

The~accretion disk can be a~thin disk with Keplerian angular velocity profile \citep{Nov-Tho:1973:BlaHol:}, or a~thick toroidal disk with angular velocity profile given by the~distribution of the~specific angular momentum of the~fluid \citep{Koz-Jar-Abr:1978:ASTRA:,Abr-Jar-Sik:1978:ASTRA:,Stu-Sla-Hle:2000:ASTRA:,Stu-Kov:2008:INTJMD:PsNewtSdS,Stu-Sla-Kov:2009:CLAQG:PseNewSdS}. The~frequency of the~oscillations is related to the~Keplerian frequency (orbital frequency of tori) and to the~radial and vertical epicyclic frequencies of the~circular test particle motion. The~epicyclic frequencies can be relevant both for the~thin, Keplerian disks with quasicircular geodetical motion \citep{Kat-Fuk-Min:1998:BHAccDis:,Klu-etal:2007:REVHA:,Kat:2001:PUBASJ:,Kat:2001:PUBASJ:b,Kat:2004:PUBASJ:QPOsmodel,Ter-Abr-Klu:2005:ASTRA:QPOresmodel,Ter-Stu:2005:ASTRA:} and for slender toroidal disks \citep{Sch-Rez:2006:ASTRJ2:QPOsRelTor,Rez-etal:2003:MNRAS:}. However, with the~thickness of an~oscillating toroid growing, the~eigenfrequencies of its radial and vertical oscillations deviate from the~epicyclic test particle frequencies \citep{Bla-etal:2006:ASTRJ2:,Str-Sra:2009:CLAQG:EpiOscNonSleKerrBH}. Here we focus our attention on the~Keplerian thin disks.

A~variety of different versions of the~orbital resonance model exists. They can be classified due to the~three following criteria:
\begin{enumerate}[a)]
    \item the~type of the~resonance (parametric or forced),
    \item the~presence of beat, combinational frequencies,
    \item the~type of oscillations entering the~resonance.
\end{enumerate}
According to the~basical criterion (a), two main groups of orbital resonance model versions exist, differing by the~type of the~resonance. In both of them, the~epicyclic frequencies of the~equatorial test particle circular motion play a~crucial role \citep{Ter-Abr-Klu:2005:ASTRA:QPOresmodel}.

\subsection{Parametric internal resonance}

The~internal resonance model assumes \emph{parametric resonance} between vertical and radial epicyclic oscillations with the~frequencies $\nu_\theta=\omega_\theta/2\pi$ and $\nu_{r}=\omega_{r}/2\pi$. The~parametric resonance is described by the~Mathieu equation \citep{Lan-Lif:1976:Mech:}
\begin{equation}
\label{Mathieu} \delta \ddot \theta + \omega_{\theta}^2\,[ 1 + h
\cos (\omega_{r} t) ]\, \delta \theta = 0.
\end{equation}
The theory behind the~Mathieu equation implies that a~parametric resonance is excited when
\begin{equation}
\label{Equation6} {\frac{\omega_{r}} {\omega_{\theta}}} =
{\frac{\nu_{r}}  {\nu_{\theta}}} = {\frac{2}  {n}}, ~~~~n
=1, \,2, \,3,\,\dots
\end{equation}
and is strongest for the~lowest possible value of $n$ \citep{Lan-Lif:1976:Mech:}. Because there is $\nu_{r} < \nu_{\theta}$ near black holes, the~lowest possible value for the~parametric resonance in the~so-called \emph{epicyclic resonance model} is $n = 3$ implying $2 \nu_{\theta} = 3 \nu_{r}$. This explains 3\,:\,2 ratio observed in the~microquasars, if $\nu_{\mathrm{U}} = \nu_{\theta}$ and $\nu_{\mathrm{L}} = \nu_{r}$. The~internal resonance corresponds to a~system with conserved energy, as shown in \citet{Hor-etal:2009:ASTRA:IntResQPOs}.

\subsection{Forced resonance}

The~\emph{forced resonance model} comes from the~idea of a~forced non-linear oscillator, when oscillations are governed by
\begin{equation}
\delta \ddot \theta + \omega_{\theta}^2\delta \theta +
\left[{\mathrm{non~linear~terms~in}}~\delta \theta \right] =
g (r)\cos (\omega_0\,t),
\end{equation}
\begin{equation}
\delta \ddot r + \omega_{{r}}^2\delta r +
\left[{\mathrm{non~linear~terms~in}}~\delta \theta, \delta r \right] =
h (r)\cos (\omega_0\,t),
\end{equation}
with
\begin{equation}
\omega_{\theta} = \left(\frac{k}{l}\right)\, \omega_{r},
\end{equation}
where $k,l$ are small natural numbers and $\omega_0$ is the~frequency of the~external force. The~non-linear terms allow combination (beat) frequencies in resonant solutions for $\delta \theta (t)$ and $\delta r(t)$ \citep[see, e.g.,][]{Lan-Lif:1976:Mech:}, which in the~simplest case give
\begin{equation}
\omega_- = \omega_{\theta} - \omega_{r}, ~~~\omega_+ =\omega_{\theta} + \omega_{r}.
\end{equation}
Such resonances can produce the~observable frequencies in the~3\,:\,2 ratio, as well as in other rational ratios. (One of the~cases that give 3\,:\,2 observed ratio is also the~``direct'' case of $k\!:\!l=3\!:\!2$ corresponding to the~same frequencies and radius as in the~case of 3\,:\,2 parametric resonance.)

The \emph{``Keplerian'' resonance} model assumes parametric or forced resonances between oscillations with radial epicyclic frequency $\nu_{r}$ and Keplerian orbital frequency $\nu_\mathrm{K}$. Of course, there are many additional possibilities for composing the~resonance conditions using the~Keplerian orbital and epicyclic frequencies in the~framework of the~multi-resonance model \citep{Stu-Kot-Tor:2013:ASTRA:MultiRez}.

The~resonance conditions related to the~frequency ratio of oscillations are common, however, physical details, such as the~time evolution of the~resonance, the~dependence of the~resonance strength and the~resonant frequency width on the~order of the~resonance, are different \citep[see][]{Lan-Lif:1976:Mech:,Nay-Moo:1979:NonOscilations:,Stu-Kot-Tor:2008:ACTA:BHadmStrResPhen}. Here we concentrate on the~frequency-ratio resonant conditions only.

\subsection{Orbital and epicyclic frequencies of the~Keplerian motion}

The~formulae for the~vertical epicyclic frequency $\nu_{\theta}$ and the~radial epicyclic frequency
$\nu_{r}$ take in the~Kerr spacetime (with the~mass $M$ and dimensionless spin $a$) the~form
\begin{equation}
\label{frequencies}
\nu_{\theta}^2 = \alpha_\theta\,\nu_\mathrm{K}^2,
\qquad
\nu_{r}^2 = \alpha_{r}\,\nu_\mathrm{K}^2
\end{equation}
\citep[e.g.,][]{Ali-Gal:1981:GENRG2:,Kat-Fuk-Min:1998:BHAccDis:,Ste-Vie:1998:ASTRJ2L:,Ter-Stu:2005:ASTRA:} where the~Keplerian frequency $\nu_\mathrm{K}$ and related epicyclic frequencies are given by the~formulae
\begin{eqnarray}
\nu_{\mathrm{K}}&=&\frac{1}{2\pi}\left(\frac{\mathrm{G}M}{r_\mathrm{G}^{~3}}\right)^{1/2}\frac{1}{x^{3/2} + a} =
\frac{1}{2\pi}\left(\frac{\mathrm{c}^3}{\mathrm{G}M}\right)
\frac{1}{x^{3/2}+a}\,,\\
\alpha_\theta&=& 1-\frac{4\,a}{x^{3/2}}+\frac{3\,a^2}{x^{2}}\,,\\
\alpha_{r}&=&1-\frac{6}{x}+\frac{8\,a}{x^{3/2}}-\frac{3\,a^2}{x^{2}}\,.
\end{eqnarray}
Here $x = r/(\mathrm{G}M/\mathrm{c}^2)$ is the~dimensionless radius, expressed in terms of the~gravitational radius of the~black hole, and $a$ is the~dimensionless spin.

The~Keplerian orbital frequency $\nu_{\mathrm{K}}(x,a)$ is a~monotonically decreasing function of the~radial coordinate for any value of the~black hole spin. The~radial epicyclic frequency has a~global maximum for any Kerr black hole. However, the~vertical epicyclic frequency is also not monotonic if the~spin is sufficiently high \citep[see, e.g.,][]{Kat-Fuk-Min:1998:BHAccDis:,Ter-Stu:2005:ASTRA:}. For the~Kerr spacetimes, the~locations $\mathcal{R}_{r}\,(a),~\mathcal{R}_\theta\,(a)$ of maxima of the~epicyclic frequencies $\nu_{r},~\nu_\theta$ are implicitly given by the~conditions \citep{Ter-Stu:2005:ASTRA:}
\begin{eqnarray}
\label{implicitcondition}
\beta_\mathrm{j}(x,a)&=&\frac{1}{2}\frac{\sqrt{x}}{x^{3/2}+a}\,\alpha_\mathrm{j}(x,a)\,,\quad \mathrm{where}\ ~\mathrm{j}\in\{{r},\theta\}\,,\\
\beta_{{r}}(x,a)&\equiv&\frac{1}{x^{2}}-\frac{2\,a}{x^{5/2}}+ \frac {a^2}
{x^3}\,,\\
\beta_{\theta}(x,a)&\equiv&\frac{a}{x^{5/2}}-\frac{a^2}{x^3}\,.
\end{eqnarray}

For any black hole spin $a$, the~extrema of the~radial epicyclic frequency $\mathcal{R}_{r}\,(a)$ must be located above the~marginally stable orbit. On the~other hand, the~latitudinal extrema $\mathcal{R}_\theta\,(a)$ are located above the~photon (marginally bound or marginally stable) circular orbit only if the~limits on the~black hole spin $a>0.748$ (0.852, 0.952) are satisfied \citep{Ter-Stu:2005:ASTRA:}. In the~Keplerian disks, with the~inner boundary $x_{\mathrm{in}} \sim x_{\mathrm{ms}}$, the~limiting value $a=0.952$ is relevant. In the~Kerr black hole spacetimes ($0<a<1$) there is always $\nu_{\mathrm{K}}(x) > \nu_{\theta}(x) > \nu_{r}(x)$, while in the~Kerr naked singularity spacetimes ($a>1$), the~situation is more complex and spin dependent \citep{Ter-Stu:2005:ASTRA:,Stu-Sch:2012:CQG:ObsPhenKerrSSp:}. 

For a~particular resonance $n:m$, the~equation
\begin{equation}
\label{ratios} n \nu_{r} = m \nu_{\mathrm{v}};
\qquad\nu_\mathrm{v}\in\{\nu_\theta,\, \nu_\mathrm{K}\}
\end{equation}
determines the~dimensionless resonance radius $x_{n:m}$ as a~function of the~spin $a$ in the~case of direct resonances. This can be easily extended to the~resonances with combinational frequencies \citep{Stu-Kot-Tor:2013:ASTRA:MultiRez}.

The known mass of the~central black hole, the~observed twin-peak frequencies ($\nu_{\mathrm{U}}$, $\nu_{\mathrm{L}}$), the~equations (\ref{frequencies})\,--\,(\ref{ratios}) and a~concrete type of resonance, assumed to be direct or to have the~combinational frequencies, enable to determine the~black hole spin. This procedure was first applied to the~microquasar GRO~1655$-$40 by \citet{Abr-Klu:2001:ASTRA:}, to the~other three microquasars \citep{Ter-Abr-Klu:2005:ASTRA:QPOresmodel}, and also to the~Galaxy centre black hole Sgr\,A$^*$ \citep{Ter:2005:ASTRA:}.

More complex resonant phenomena in HF~QPOs have to be expected in the~field of Kerr naked  singularities \citep{Ter-Stu:2005:ASTRA:,Stu-Sch:2012:CQG:ObsPhenKerrSSp:}. In the~naked singularity backgrounds, the~optical effects also demonstrate considerable differences as compared with those generated in the~field of black holes \citep{Stu-Sch:2010:CLAQG:AppKepDiOrKerrSSp,Stu-Sch:2012:CQG:CountRotKerrSSp:,Stu-Sch:2012:CQG:ObsPhenKerrSSp:,Vir-Ell:2002:,Vir-Kee:2008:PHYSR4:TimDelGrLens,Tak-Har:2010:CLAQG:ObsTestKerr,Eir:2012:arXiv}. Here we restrict our attention to the~black hole spacetimes.

\section{Multi-resonance model}

The~very probable interpretation of twin peak frequencies observed in microquasars is the~3\,:\,2 parametric resonance of the~epicyclic oscillations; however, we expect that more than one resonance could be excited in the~disk simultaneously (or at different times) under different internal conditions. Indeed, observations of the~HF~QPOs in the~microquasar GRS~1915$+$105 \citep{Rem:2005:ASTRN:}, in the~extragalactic sources NGC 4051, MCG-6-30-15 \citep[]{Lac-Cze-Abr:2006:astro-ph0607594:}, NGC~5408~X-1 \citep[]{Str-etal:2007:ASTRJ2:QuaPerVar}, and the~Galaxy centre Sgr\,A$^*$ \citep[]{Asc-etal:2004:ASTRA:} show a~variety of QPOs with frequency ratios differing from (or additional to) the~3\,:\,2 ratio.

In the~framework of the~multi-resonance model we consider two different resonances determined by a~doubled ratio of natural numbers $n$\,:\,$m$ and $n'$\,:\,$m'$. They occur at corresponding radii of the~disk $x_{n:m}$, $x_{n':m'}$ that can be determined from the~observed set of frequencies using the~relevant versions of the~orbital resonance model. In a~degenerated case when only triple frequency set is observed, we can distinguish in the~multi-resonance model two relevant cases. First, the~three frequencies could be related to three oscillating modes occurring at a~common radius, if the~black hole has a~specific, ``magic'', spin given by the~frequency ratio of the~set -- this case corresponds to the~strong resonance model as we can assume cooperative resonant phenomena occurring at the~common radius \citep{Stu-Kot-Tor:2008:ACTA:BHadmStrResPhen,Stu-Kot-Tor:2013:ASTRA:MultiRez}. Second, three frequency sets with a~duplex frequency could occur accidentally for proper values of the~black hole dimensionless spin. In such a~case, two twin peak QPOs observed at the~radii $x_{n:m}$ and $x_{n':m'}$ have the~bottom, top, or mixed (the~bottom at the~inner radius and the~top in the~outer radius, or vice versa) frequencies identical. Such situations can be characterized by sets of three frequencies (upper $\nu_{\mathrm{U}}$, middle $\nu_{\mathrm{M}}$ and lower $\nu_{\mathrm{L}}$) with ratio $\nu_{\mathrm{U}}:\nu_{\mathrm{M}}:\nu_{\mathrm{L}} = s:t:u$, given by the~$n:m$ and $n':m'$ ratios, the~relevant versions of the~resonance, and the~type of the~duplex (common) frequency. The guide book of the~possible combinations generated by the~orbital and epicyclic frequencies and their combinations is presented in \citet{Stu-Kot-Tor:2013:ASTRA:MultiRez}. When only direct resonances of the~epicyclic oscillations are allowed, the~first case with ``bottom identity'' can be realized by the~situation with two resonances having common radial epicyclic frequency, while the~second case with ``top identity'' can be realized by the~situation with two resonances having common vertical epicyclic frequency. These two possibilities only are in principle allowed by the~non-monotonicity of the~epicyclic frequencies (\ref{frequencies}) discussed in detail by \citet{Tor-Stu:2005:RAGtime6and7:CrossRef,Ter-Stu:2005:ASTRA:}. When the~Keplerian oscillations and the~combinational frequencies are allowed, all the~mixed, bottom, and top identities are possible. We consider frequency ratios of small integers, with the~order of the~resonances $n+m \leq 9$ ($n'+m' \leq 9$), since the~resonant phenomena are realistic only for $n \leq 5$ \citep[see][for details]{Lan-Lif:1976:Mech:,Nay-Moo:1979:NonOscilations:}.

\subsection{Resonance conditions}

For all possible resonances of the~epicyclic and Keplerian frequencies, the~resonance condition for the~ratio $\nu_{\mathrm{U}} : \nu_{\mathrm{L}} = n:m$ is given in terms of the~frequency ratio parameter
\begin{equation}
p=\left(\frac{m}{n}\right)^2.
\end{equation}
The~resonant conditions that implicitly determine the~resonant radius $x_{n:m}(a,p)$ have to be related to the~radius of the~innermost stable circular geodesic $x_{\mathrm{ms}}(a)$ giving the~inner edge of Keplerian disks. We require $x_{n:m}(a,p) \geq x_{\mathrm{ms}}(a)$, where $x_{\mathrm{ms}}(a)$ is implicitly given by
\begin{equation}
a=a_{\mathrm{ms}}\equiv\frac{\sqrt{x}}{3}\left(4-\sqrt{3x-2}\right).
\end{equation}
The resonance functions are denoted as $a^{\nu_{\mathrm{U}}/\nu_{\mathrm{L}}}(x,p)$.  As an~example we give the~resonance function (and the~resonance condition) for the~radial and vertical epicyclic frequencies ($\nu_{\mathrm{U}} = \nu_{\theta},\nu_{\mathrm{L}} = \nu_{r}$) that read
\begin{equation}\label{aD1}
          a=a^{\theta/{r}}(x,p)\equiv\frac{\sqrt{x}}{3(p+1)}\Bigg\{2(p+2)-\sqrt{(1-p)\left[3x(p+1)-2(2p+1)\right]}\Bigg\}\,,
\end{equation}
All the~other resonance conditions and functions for direct and combinational resonances can be found in \citet{Stu-Kot-Tor:2013:ASTRA:MultiRez}.

\subsection{Frequency sets with a~duplex frequency}

From the~point of view of the~observational consequences, it is important to know for which frequency ratios $n\!:\!m$ the~resonant frequency $\nu_{\theta}(a,n\!:\!m)$, which is considered as a~function of the~black hole spin $a$ for a~given frequency ratio $n\!:\!m$, has a~non-monotonic character. A~detailed analysis \citep[]{Ter-Stu:2005:ASTRA:} shows that $\nu_{\theta}(a,n\!:\!m)$ has a~local maximum for $n\!:\!m > 11\!:\!5$; i.e., in physically relevant situations ($n$, $m$ small enough for the~resonance), it occurs for the~ratios $\nu_{\theta}:\nu_{r}=$ 5\,:\,2, 3\,:\,1, 4\,:\,1, 5\,:\,1. This means that while the~``bottom identity'' could happen for any black hole spin $a$, the~``top identity'' can only arise for $a\sim1$ if only the~epicyclic oscillations are considered. For details see \citet{Stu-Kot-Tor:2013:ASTRA:MultiRez} -- we adopt here the~system of notation introduced in this paper: $\mathrm{T}(X)(Y)$, $\mathrm{B}(X)(Y)$, and $\mathrm{M}(X)(Y)$ where T stands for top, B for bottom, and M for mixed identity. The~$(X)(Y)$ corresponds to the~given types of the~doubled resonances $(X)$ and $(Y)$ with identical top, bottom, or mixed frequencies (for the~case of mixed identity $(X)$ always denotes the~resonance with the~top frequency and $(Y)$ the~resonance with the~identical bottom frequency). For the~triple frequency sets with the~ratio $3:2:1$, the~concrete types of the~resonances $(X)(Y)$ are presented in Table~\ref{relace}.

\renewcommand{\arraystretch}{1.4}
\begin{table}[!h]
\caption{{\small \label{relace}Relevant versions of the~multi-resonant model with assumed observed characteristic frequency ratio set $\nu_{\mathrm{U}}:\nu_{\mathrm{M}}:\nu_{\mathrm{L}}=$ 3\,:\,2\,:\,1.}}
\begin{center}
\begin{tabular}{lccc}
 \hline
    Set & {\rule[-2mm]{0mm}{6mm}
$\nu_{\mathrm{U}}$} & $\nu_{\mathrm{M}}$ & $\nu_{\mathrm{L}}$ \\
 \hline
T23 (``magic'')  & $\nu_{\mathrm{K}}$ & $\nu_{\theta}^{3:2}$ &
$\nu_{r}^{3:1}$   \\
T45 & $\nu_{\mathrm{K}}^{3:1}=\nu_{\theta}^{3:2}$ &
$\left(\nu_{\mathrm{K}}-\nu_{r}\right)^{3:2}$ &
$\left(\nu_{\theta}-\nu_{r}\right)^{3:1}$ \\
B15 & $\nu_{\theta}^{3:1}$ & $\nu_{\theta}^{2:1}$ &
$\nu_{r}^{3:1}=\left(\nu_{\mathrm{K}}-\nu_{r}\right)^{2:1}$ \\
 B22 & $\nu_{\mathrm{K}}^{3:1}$ & $\nu_{\mathrm{K}}^{2:1}$ & $\nu_{r}$ \\
 B24 & $\nu_{\mathrm{K}}^{3:1}$ & $\nu_{\mathrm{K}}^{2:1}$ & $\nu_{r}^{2:1}=\left(\nu_{\theta}-\nu_{r}\right)^{3:1}$ \\
 B25 & $\nu_{\mathrm{K}}^{3:1}$ & $\nu_{\theta}^{2:1}$ & $\nu_{r}^{3:1}=\left(\nu_{\mathrm{K}}-\nu_{r}\right)^{2:1}$ \\
 B45 & $\nu_{\mathrm{K}}^{3:1}$ & $\nu_{\theta}^{2:1}$ & $\left(\nu_{\theta}-\nu_{r}\right)^{3:1}=\left(\nu_{\mathrm{K}}-\nu_{r}\right)^{2:1}$ \\
 M14 & $\nu_{\mathrm{K}}^{3:2}$ & $\nu_{\theta}^{2:1}=\left(\nu_{\theta}-\nu_{r}\right)^{3:2}$ & $\nu_{r}^{2:1}$ \\
 M24 & $\nu_{\mathrm{K}}^{3:2}$ & $\nu_{\mathrm{K}}^{2:1}=\left(\nu_{\theta}-\nu_{r}\right)^{3:2}$ & $\nu_{r}^{2:1}$ \\
 M54 & $\nu_{\mathrm{K}}^{3:2}$ & $\nu_{\theta}^{2:1}=\left(\nu_{\theta}-\nu_{r}\right)^{3:2}$ & $\left(\nu_{\mathrm{K}}-\nu_{r}\right)^{2:1}$ \\
T1(DC1) & $\left(\nu_{\mathrm{K}}-\nu_{r}\right)^{3:2}=\nu_{\theta}^{3:1}$ &
   $\left(\nu_{\theta}-\nu_{r}\right)^{3:2}$ & $\nu_{r}^{3:1}$ \\
 B1(DC11)
 & $\nu_{\theta}^{3:1}$ &
  $\left(\nu_{\theta}-\nu_{r}\right)^{2:1}$ &
  $\nu_{r}^{3:1}=\left(\nu_{\mathrm{K}}-\nu_{\theta}\right)^{2:1}$ \\
 T1(DC3)
 & $\left(\nu_{\mathrm{K}}-\nu_{\theta}\right)^{3:2}=\nu_{\theta}^{3:1}$ &
  $\left(\nu_{\theta}-\nu_{r}\right)^{3:2}$ &
  $\nu_{r}^{3:1}$ \\
 T1(DC9)
 &  $\left(\nu_{\mathrm{K}}-\nu_{r}\right)^{3:1}=\nu_{\theta}^{3:2}$ &
  $\nu_{r}^{3:2}$ &
  $\left(\nu_{\mathrm{K}}-\nu_{\theta}\right)^{3:1}$ \\
 M(DC11)
 & $\nu_{\theta}^{3:2}$ &
  $\nu_{r}^{3:2}=\left(\nu_{\theta}-\nu_{r}\right)^{2:1}$ &
  $\left(\nu_{\mathrm{K}}-\nu_{\theta}\right)^{2:1}$ \\
 T1(DC11)$^\mathrm{a}$
 & $\left(\nu_{\theta}-\nu_{r}\right)^{3:1}=\nu_{\theta}^{3:2}$ &
  $\nu_{r}^{3:2}$ &
  $\left(\nu_{\mathrm{K}}-\nu_{\theta}\right)^{3:1}$ \\
 T1(DC11)$^\mathrm{b}$
 & $\left(\nu_{\theta}-\nu_{r}\right)^{3:2}=\nu_{\theta}^{3:1}$ &
  $\left(\nu_{\mathrm{K}}-\nu_{\theta}\right)^{3:2}$ &
  $\nu_{r}^{3:1}$ \\
 B1(DC9)
 & $\left(\nu_{\mathrm{K}}-\nu_{r}\right)^{3:1}$ &
  $\nu_{\theta}^{2:1}$ &
  $\nu_{r}^{2:1}=\left(\nu_{\mathrm{K}}-\nu_{\theta}\right)^{3:1}$ \\
 M1(DC1)
 & $\left(\nu_{\mathrm{K}}-\nu_{r}\right)^{3:2}$ &
  $\nu_{\theta}^{2:1}=\left(\nu_{\theta}-\nu_{r}\right)^{3:2}$ &
  $\nu_{r}^{2:1}$ \\
 B1(DC12)
 & $\left(\nu_{\theta}+\nu_{r}\right)^{3:1}$ &
  $\nu_{\theta}^{2:1}$ &
  $\nu_{r}^{2:1}=\left(\nu_{\mathrm{K}}-\nu_{\theta}\right)^{3:1}$ \\
 M1(DC7)
 & $\left(\nu_{\mathrm{K}}-\nu_{\theta}\right)^{3:2}$ &
  $\nu_{\theta}^{2:1}=\left(\nu_{\theta}+\nu_{r}\right)^{3:2}$ &
  $\nu_{r}^{2:1}$ \\
 \hline
\end{tabular}
\end{center}
\end{table}

\subsection{Triple frequency sets and black hole spin}

The multi-resonance model gives strong restrictions on the~black hole spin $a$ when the~triple frequency sets are considered \citep{Stu-Kot-Tor:2013:ASTRA:MultiRez}. For the~simple case of the~``top identity'' of the~upper frequencies in two resonances between the~radial and vertical epicyclic oscillations at the~radii $x_{p}, x_{p'}$ with $p^{1/2}=m:n$, $p'^{1/2}=m':n'$, the~condition
$\nu_{\theta}(a,x_p)=\nu_{\theta}(a,x_{p'})$ can be transformed to the~relation
\begin{equation}
\frac{\alpha_{\theta}^{1/2}(a,x_{p})}{x_{p}^{3/2}+a} =
\frac{\alpha_{\theta}^{1/2}(a,x_{p'})}{x_{p'}^{3/2}+a}\,,
\end{equation}
which uniquely determines the~black hole spin $a$. When two different resonances are combined, we proceed in the~same manner. For example, the~case of ``bottom identity'' in the~resonance between the~radial and vertical epicyclic oscillations at $x_{p}$ and the~resonance between the~oscillations with the~Keplerian frequency $\nu_{\mathrm{K}}$ and total precession oscillations with $\nu_{\mathrm{T}}=\nu_{\theta}-\nu_{r}$ at $x_{p'}$ implies the~condition
$\nu_{r}(a,x_p)=\left(\nu_{\theta}-\nu_{r}\right)(a,x_{p'})$ that leads to the~relation
\begin{equation}
\frac{\alpha_{{r}}^{1/2}(a,x_{p})}{x_{p}^{3/2}+a} =
\frac{\left(\alpha_{\theta}-\alpha_{{r}}\right)(a,x_{p'})}{x_{p'}^{3/2}+a}\,,
\end{equation}
which uniquely determines the~dimensionless spin $a$, since the~radii $x_{p}$ and $x_{p'}$ are related to the~spin $a$ by the~resonance conditions for $a^{\theta/{r}}(x,p)$ (see Eq.~(\ref{aD1})) and $a^{\mathrm{K}/(\theta-{r})}(x,p')$, respectively. Therefore, for given types of the~doubled resonances, the~ratios $n:m$ and $n':m'$ determine the~ratio in the~triple frequency set $s:t:u$. The~black hole spin $a$ is given by the~types of the~two resonances and the~ratios $p, p'$, quite independently of the~black hole mass $M$.

Since the~radial and vertical epicyclic frequencies and the~orbital frequency have the~same dependence on the~black hole mass $M$, the~above arguments hold for any kind of the~three frequency sets, for any of the~bottom, top, or mixed frequency identity with any two resonances containing any combination of the~frequencies $\nu_{\mathrm{K}}, \nu_{\theta}, \nu_{r}$. Therefore, the~triple frequency sets with the~``duplex'' frequencies can be used to determine the~black hole spin with very high precision given by precision of the~frequency measurements that can be very high \citep{Stu-Kot-Tor:2013:ASTRA:MultiRez}. The~mass parameter $M$ can be addressed by the~magnitude of the~measured frequencies. However, the~relation between the~black hole spin and the~triple frequency ratios is not unique in general. For a~given frequency ratio set, several values of $a$ are allowed, and some other methods of the~spin measurement (spectral continuum fitting, profiled spectral lines) must be involved. It is then important that the~spin measurements can be considered quite independently of the~mass measurements based on different methods.

\subsection{Strong resonant radii and related black hole spin}

In situations discussed above, the~triple frequency sets fixing the~black hole spin $a$ occur at two different radii. However, there are important cases when the~triple frequencies occur at the~same (shared) radius \citep{Stu-Kot-Tor:2008:ACTA:BHadmStrResPhen}. Then we expect higher probability that the~resonant phenomena will arise, especially in the~cases of ratios of very low integers because a~causally related cooperation of the~resonances at the~given radius can be relevant. A~crucial role is expected for direct resonances of oscillations with all three orbital frequencies characterized by a~triple frequency ratio set ($s$, $t$, $u$ being small natural numbers)
\begin{equation}
\nu_{\mathrm{K}} : \nu_{\theta} : \nu_{r} = s:t:u
\end{equation}
when strong resonant phenomena are possible. The~radius giving strong resonance $s:t:u$ ratio is given by \citep{Stu-Kot-Tor:2008:ACTA:BHadmStrResPhen}
\begin{eqnarray}
&&x\left(s/u,t/u\right)\equiv 6\left(s/u\right)^2{X}^{-1}\,,\\
&&X=6 \left(s/u\right)^2\pm 2 \sqrt{2} \sqrt{\left(t/u-1\right)
\left(t/u+1\right) \left[3 \left(s/u\right)^2-\left(t/u\right)^2-2
\right]} - \left[\left(t/u\right)^2+5\right]\,,\nonumber
    \end{eqnarray}
and the~related black hole ``magic'' spin is given by
   \begin{equation}
a\left(x\left(s/u,t/u\right),u/s\right)\equiv\frac{\sqrt{x}}{3}\left(4\pm\sqrt{-2+3x\left[1-\left(u/s\right)^2\right]}\,\right).
   \end{equation}
A~detailed discussion of the~black holes that admit strong resonant phenomena can be found for small integers ($s\leq 5$) in \citet{Stu-Kot-Tor:2008:ACTA:BHadmStrResPhen}.

Of special interest is the~case of the~``magic'' spin $a=0.983$, when the~Keplerian and epicyclic frequencies are in the~ratio $\nu_{\mathrm{K}}:\nu_{\theta}:\nu_{r} = 3:2:1$ at the~common radius $x_{3:2:1} = 2.395$. In fact, this case involves rather extended structure of resonances \citep[see][]{Stu-Kot-Tor:2008:ACTA:BHadmStrResPhen}. It should be stressed that beside the~case of strong resonances between oscillations with $\nu_{\mathrm{K}}$, $\nu_{\theta}$, $\nu_{r}$ sharing the~same radius, the~characteristic set 3\,:\,2\,:\,1 could appear also due to resonances at different radii \citep[see Fig.~5 in][]{Stu-Kot-Tor:2013:ASTRA:MultiRez}. All the~relevant versions of the~multi-resonant model with 3\,:\,2\,:\,1 frequency ratio set are given in Table~\ref{relace}.

\section{Acceptable variants of the~multi-resonance orbital model and possible spin of~the~supermassive black hole at~Sgr\,A$^*$}

\subsection{HF~QPOs observed in Sgr\,A$^*$}
The Galaxy centre supermassive black hole source Sgr\,A$^*$ serves as an~appropriate astrophysical test system for the~multi-resonance orbital model. The~HF~QPOs with frequency ratio $\sim 3\!:\!2\!:\!1$ were reported for Sgr\,A$^*$ \citep[][]{Asch:2004:ASTRA:,Asc-etal:2004:ASTRA:,Ter:2005:ASTRA:} and represent therefore an~appropriately simple sample to test the~multi-resonant model. Although there are doubts on validity of the~data that are not fully accepted by the~astrophysical community, we feel it could be important, illustrative and interesting to test possible implications of the~observations, assuming their relevance.

The observed frequency ratio is
 \begin{equation}
              \frac{1}{692} : \frac{1}{1130} : \frac{1}{2178} \sim 3 : 2 : 1 .
 \end{equation}
The upper frequency was observed with a~rather high error \citep{Asch:2004:ASTRA:}
\begin{equation}
               \nu_{\mathrm{U}} = (1.445 \pm 0.16)\,\mathrm{mHz} .
\end{equation}
Although the~reported HF~QPOs are considered as controversial, it is still worth to consider them seriously and take them as a~starting point to demonstrate the~effectiveness of the~multi-resonance model in the~most interesting case of triple frequency set with the~ratio $3:2:1$. The frequency measurement precision is very poor in the~case of data given in \citet{Asch:2004:ASTRA:}, it is much worse in comparison with data obtained by measurements in the~LMXBs containing microquasars, where the~precision is of the~order of $1\%$ \citep{Rem-McCli:2006:ARASTRA:,McCli-etal:2011:CLAQG:} and can be expected by one order better in the~planned X-ray satellite observatory LOFT \citep{Fer-etal:2012:ExA:}. Here we assume for simplicity the~frequency ratio condition $3:2:1$ to be exactly fulfilled, giving thus the~spin precisely, and shifting all the~uncertainties of the~frequency measurement to the~estimates of the~black hole mass that is given by the~magnitude of the~measured frequency.\footnote{Error in determination of the~dimensionless spin of the~black hole is discussed in \citet{Stu-Kot-Tor:2008:ACTA:BHadmStrResPhen,Stu-Kot-Tor:2013:ASTRA:MultiRez}.}

Comparing the~mass uncertainty related to the~measurement error of the~upper frequency to the~range of mass implied by the~motion of the~stars observed in vicinity of the~Sgr\,A$^*$ \citep{Ghe-etal:2008:SgrA:}, we are able to find acceptable variants of the~multi-resonance model predicting the~related spin of the~supermassive black hole at Sgr\,A$^*$.

\subsection{Sgr\,A$^*$ black hole parameters}

\begin{figure}[t]
\begin{center}
\includegraphics[width=.9\hsize]{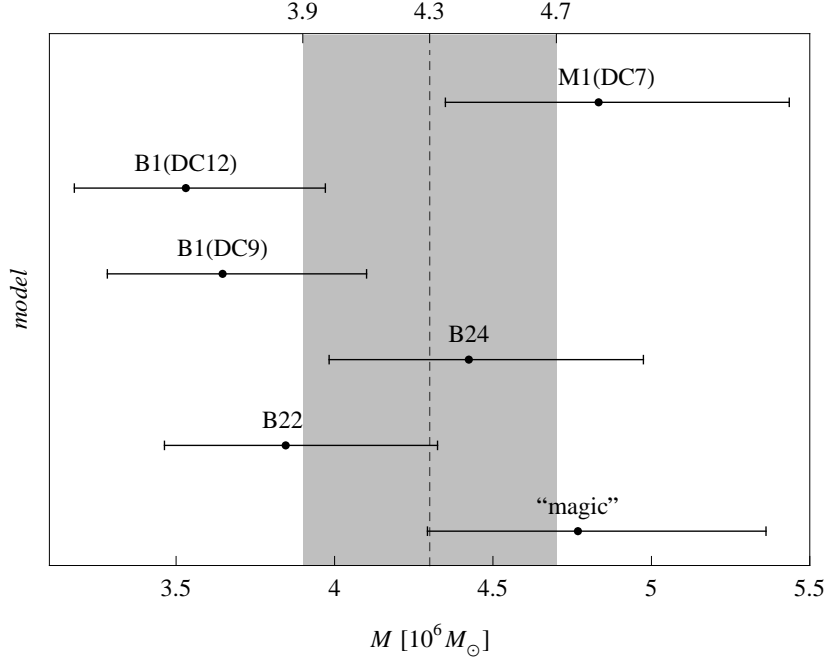}
\end{center}
\caption{\small Mass of Sgr\,A$^*$: six versions of multi-resonance model (see Table~\ref{Sgr-vysledky}) that are compatible with mass estimates given by the~star orbital motion
\citep{Gil-etal:2009:Mass-SgrA:}, illustrated here by the~grey rectangle.\label{offsety}}
\end{figure}

It has been shown that if the~strong resonant model with $\nu_{\mathrm{K}}:\nu_{\theta}:\nu_{r} =$ 3\,:\,2\,:\,1 is applied \citep{Stu-Kot-Tor:2008:ACTA:BHadmStrResPhen}, the~observed data imply spin $a$ and mass $M$ of Sgr\,A$^*$ black hole not contradicting the~black hole mass estimate given by the~star orbital motion that reads $M_{\mathrm{Sgr}} \sim 4.3 \times 10^{6}\,\mathrm{M}_{\odot}$ \citep{Gil-etal:2009:Mass-SgrA:}
with the~interval of allowed values given by
\begin{equation}
    3.9 \times 10^6\,\mathrm{M}_{\odot} < M < 4.7 \times 10^6\,\mathrm{M}_{\odot}\,,
    \end{equation}
considering also the~error given by uncertainty in distance measurement to Sgr\,A$^*$. Here, we test all the~relevant versions of the~multi-resonant orbital model. The~results are summarized in Fig.~\ref{offsety} and related Table~\ref{Sgr-vysledky}, including the~case of the~strong resonant phenomena.

\renewcommand{\arraystretch}{1.4}
\begin{table}[!h]
\caption{{\small \label{Sgr-vysledky}Black hole spin and mass of Sgr\,A$^*$ calculated for all the~relevant versions of the~multi-resonant model with assumed observed characteristic frequency ratio set $\nu_{\mathrm{U}}:\nu_{\mathrm{M}}:\nu_{\mathrm{L}}=$
3\,:\,2\,:\,1; $\nu_{\mathrm{U}}=(1.445\pm 0.16)\,\mathrm{mHz}$
is used to determine the~black hole mass. The~radius of marginally stable orbit $x_\mathrm{ms}$ and corresponding resonant radii $x_\mathrm{1}$ and $x_\mathrm{2}$ are given. An~asterisk denotes the~special values of the~black hole spin when the~resonance points share the~same radius ($x_\mathrm{1} = x_\mathrm{2}\equiv x_\mathrm{3:2:1}$). Gray shaded rows represent the~resonant model versions that are compatible with mass estimates given by the~star orbital motion \citep{Gil-etal:2009:Mass-SgrA:}. Dark grey shaded row corresponds to the~best fit, see also Fig.~\ref{offsety}.}}
\begin{center}
\begin{tabular}{lccccc}
 \hline
    Set & {\rule[-2mm]{0mm}{6mm}$a$} & $x_{\mathrm{ms}}$ & $x_1$ & $x_2$ & $M\,\left[10^6\,\mathrm{M}_{\odot}\right]$\\
 \hline
\LCC \lg & \lg & \lg & \lg & \lg & \lg\\
T23 (``magic'')  & $0.983043$* &
    $1.571$ & \multicolumn{2}{c}{$2.395$} & $4.293-5.362$
    \\
    \ECC
T45 & $0.885010${\phantom{*}} &
    $2.419$ & $3.720$ & $4.299$ & $2.054-2.566$ \\
B15 & $0.616894${\phantom{*}} & $3.758$ & $4.241$ & $5.833$ & $1.903-2.376$ \\
B15 & $0.999667${\phantom{*}} & $1.121$ & $1.411$ & $4.250$ & $2.606-3.255$
 \\
\LCC \lg & \lg & \lg & \lg & \lg
& \lg\\
 B22 & $0.913806${\phantom{*}} &
    $2.225$ & $2.885$ & $3.935$ & $3.463-4.325$
 \\\ECC
\LCC \mg & \mg & \mg & \mg & \mg
& \mg\\
 B24 & $0.980124${\phantom{*}} &
    $1.612$ & $2.551$ & $3.519$ & $3.983-4.975$ \\\ECC
 B25 & $0.475159${\phantom{*}} &
    $4.330$ & $4.988$ & $6.359$ & $1.733-2.165$ \\
 B45 & $0.922985${\phantom{*}}  &
    $2.158$ & $3.794$ & $4.594$ & $2.422-3.025$
    \\
 M14 & $0.544870${\phantom{*}} &
    $4.055$ & $4.347$ & $5.477$ & $2.095-2.617$ \\
 M24 & $0.535413${\phantom{*}}  &
    $4.093$ & $4.394$ & $5.832$ & $2.065-2.580$ \\
 M54 & $0.336030${\phantom{*}} &
    $4.849$ & $5.327$ & $6.857$ & $1.594-1.991$ \\
T1(DC1) & & & \multicolumn{2}{c}{} & \\
B1(DC11)
 &
 \raisebox{1.6ex}[0pt]
 {$\llap{{\Bigg\}\,\,}} 0.865670$*} &
 \raisebox{1.6ex}[0pt]
 {$2.539$} &
 \multicolumn{2}{c}
 {\raisebox{1.6ex}[0pt]
 {$2.880$}} &
 \raisebox{1.6ex}[0pt]
 {$2.625-3.278$} 
 \\
 T1(DC3) & $0.986666$* &
 $1.514$ & \multicolumn{2}{c}{$1.753$} &
 $3.043-3.800$ 
 \\
 T1(DC9) & & & & & \\
 M(DC11) &
  \raisebox{1.6ex}[0pt]
 {$\llap{{\Bigg\}\,\,}} 0.892290${\phantom{*}}} &
 \raisebox{1.6ex}[0pt]
 {$2.372$} &
 \raisebox{1.6ex}[0pt]
 {$3.601$} &
 \raisebox{1.6ex}[0pt]
 {$5.034$} &
 \raisebox{1.6ex}[0pt]
 {$1.457-1.820$} 
 \\
 T1(DC11)$^\mathrm{a}$ &  $0.772687${\phantom{*}} &
 $3.046$ & $3.792$ & $6.036$ &
 $1.183-1.477$ 
 \\
 T1(DC11)$^\mathrm{b}$ &
 $0.927324${\phantom{*}} &
 $2.125$ & $2.131$ & $2.419$ &
 $2.895-3.616$ 
 \\
\LCC \lg & \lg & \lg & \lg & \lg
& \lg\\
 B1(DC9) & & & & & \\
  \ECC
\LCC \lg & \lg & \lg & \lg & \lg
& \lg\\
 M1(DC1) &
  \raisebox{1.6ex}[0pt]
 {$\llap{{\Bigg\}\,\,}} 0.868917${\phantom{*}}} &
 \raisebox{1.6ex}[0pt]
 {$2.520$} &
 \raisebox{1.6ex}[0pt]
 {$2.648$} &
 \raisebox{1.6ex}[0pt]
 {$3.510$} &
 \raisebox{1.6ex}[0pt]
 {$3.283-4.101$} 
 \\
  \ECC
\LCC \lg & \lg & \lg & \lg & \lg
& \lg\\
 B1(DC12)
 & $0.851581${\phantom{*}} &
 $2.623$ & $2.675$ & $3.640$ &
 $3.179-3.971$ 
 \\
  \ECC
\LCC \lg & \lg & \lg & \lg & \lg
& \lg\\
 M1(DC7) &  $0.987594${\phantom{*}} &
 $1.498$ & $1.502$ & $2.326$ &
 $4.352-5.435$ 
 \\
 \ECC
 \hline
\end{tabular}
\end{center}
\end{table}

We can see in Table~\ref{Sgr-vysledky} that from all of the~theoretically allowed (19) versions, only six versions are compatible with observational restrictions from the~orbital motion of the~stars in vicinity of Sgr\,A$^*$. In all of the~six cases, the~black hole spin $a > 0.85$, in agreement with the~assumption that the~Galactic centre black hole should be fast rotating. The~best fit (B24) is obtained for the~spin $a=0.980$, and mass $M\in(3.983-4.975)\times 10^6\,\mathrm{M}_{\odot}$, with resonances $\nu_{\mathrm{K}}\!:\!\left(\nu_{\theta}-\nu_{r}\right) =$ 3\,:\,1, $\nu_{\mathrm{K}}\!:\!\nu_{r} =$ 2\,:\,1, having a~common bottom frequency. As we can see from Fig.~\ref{offsety}, for the~case B24 with $a=0.980$, the~mean value of estimated mass $M = 4.42 \times 10^{6}\,\mathrm{M}_{\odot}$ is close to the~orbital motion estimate, while in the~other five acceptable cases, the~difference is much greater, however, two of them are still well acceptable. The~good fits are obtained also for the~cases T23 and B22, where the~strong resonance for the~``magic'' spin $a = 0.983$ (T23) implies $M\in(4.293-5.362)\times 10^6\,\mathrm{M}_{\odot}$, and the~case B22 with $a=0.914$ gives the~black hole mass $M\in(3.463-4.325)\times 10^6\,\mathrm{M}_{\odot}$. The frequency relations corresponding to the~cases T23 and B22 are illustrated in Fig.~5 in \citet{Stu-Kot-Tor:2013:ASTRA:MultiRez}.

The~multi-resonance orbital model should be further tested and more precise frequency measurements are very important. We have to compare the~results of our model to the~other methods of black hole spin measurements, namely those related to the~optical phenomena in close vicinity of the~supermassive black hole at Sgr\,A$^*$ enabling to observe even silhouette of the~black hole \citep{Sch-Stu:2009:INTJMD:OpPheBraKerr,Bin-Nun:2010:PRD:SgrA:,Vir-Ell:2002:,Vir-Kee:2008:PHYSR4:TimDelGrLens,Stu-Sch:2010:CLAQG:AppKepDiOrKerrSSp,Ama-Eir:2012:PhRvD,Eir:2012:arXiv}. For Sgr\,A$^*$, the~relativistic precession of the~nearby star orbits is also very promising \citep[]{Kra:2005:,Kra:2007:}.

\section{Conclusions}

We have demonstrated that the~multi-resonant model of HF~QPOs based on the~orbital motion is capable to explain the~HF~QPO data observed in the~supermassive Galactic centre (Sgr\,A$^*$) black hole \citep{Asch:2004:ASTRA:}.

There are three versions of the~multi-resonance orbital model predicting the~mass parameter $M$ of the~central supermassive black hole at Sgr\,A$^*$ in agreement with the~restrictions determined from the~measurements of the~motion of the~stars orbiting the~black hole. All the~versions also predict relatively large spin ($a>0.9$) in accord with expectation of large spin of the~central black hole. Among the~acceptable versions the~strong resonance model with the~magic spin $a=0.983$ belongs, but the~mean value of the~Sgr\,A$^*$ black hole mass predicted by this version does not meet the~range of the~mass predicted by the~star motion. On the~other hand, the~best version (B24) predicts $a=0.980$ and  the~mean value of the~mass parameter very close to the~mean value determined by the~orbits of stars. The strong limit on the~supermassive black hole predicted by the~acceptable versions of the~multi-resonance model can be useful in studies of the~optical phenomena expected to be observed in very close vicinity of the~black hole horizon because of development of the~recent VLBI observational technique \citep{Doe-etal:2009:ApJ}.

Despite the~fact that there are doubts on the~observational data used for our study of the~HF~QPOs in the~Sgr\,A$^*$ source, we can conclude that the~multi-resonance model can work efficiently in restricting the~black hole mass parameters, especially when more precise observations of the~HF~QPOs will be obtained in the~measurements of HF~QPOs in microquasars and in galaxy centers. We expect very precise measurements obtained by the~planned X-ray satellite observatory LOFT \citep{Fer-etal:2012:SPIE,Fer-etal:2012:ExA:}. In the~case of Sgr\,A$^*$ some active accretion periods are necessary in order to have possibility of creating observable HF~QPOs.

\sloppy
\vspace{3ex}
\noindent
{\small{\bf{Acknowledgements.}}
We would like to express our gratitude to the~Czech grant GA\v{C}R~202/09/0772 and the~internal grants of the~Silesian University in Opava SGS/11/2013 and SGS/23/2013. The~authors further acknowledge the~project Supporting Integration with the~International Theoretical and Observational Research Network in Relativistic Astrophysics of Compact Objects, CZ.1.07/2.3.00/20.0071, supported by Operational Programme \emph{Education for Competitiveness} funded by Structural Funds of the~European Union and the~state budget of the~Czech Republic.}

\providecommand{\uv}[1]{\glqq#1\grqq}

\end{document}